\documentclass{article}

\usepackage{fancyheadings}

\newtheorem{example}{Example}

\title{Open Network Handles Implemented in DNS\\[2ex]
\texttt{<draft-odonnell-onhs-imp-dns-00.txt>}}

\begin{document}

\maketitle

\paragraph{Archived copy of Internet Draft\\}

This report is a hand translation into \LaTeX\ format of an Internet
Draft, published as \texttt{draft-odonnell-onhs-imp-dns-00.txt} in
plain text generated from ROFF source by the \emph{Internet
  Engineering Task Force} (IETF).  All text besides this note comes
from that Internet Draft. IETF does not keep a permanent archive of
Internet Drafts, so this report serves as a more permanent record of
the Internet Draft. It does not reproduce the precise format nor
pagination of the IETF copy.

\paragraph{Status of this Memo\\}

This document is an Internet-Draft and is subject to all provisions of
Section 10 of RFC2026.

The key words "must", "must not", "required", "shall", "shall not",
"should", "should", "recommended", "may", and "optional" in this
document are to be interpreted as described in RFC-2119.

This memo provides information for the Internet community. This memo
does not specify an Internet standard of any kind. Distribution of
this memo is unlimited.

Internet-Drafts are working documents of the Internet Engineering Task
Force (IETF), its areas, and its working groups.  Note that other
groups may also distribute working documents as Internet-Drafts.

Internet-Drafts are draft documents valid for a maximum of six months
and may be updated, replaced, or obsoleted by other documents at any
time.  It is inappropriate to use Internet- Drafts as reference
material or to cite them other than as "work in progress."

The list of current Internet-Drafts can be accessed at\\
\texttt{http://www.ietf.org/1id-abstracts.html}

The list of Internet-Draft Shadow Directories can be accessed at\\
\texttt{http://www.ietf.org/shadow.html}

\begin{abstract}

An Open Network Handle System (ONHS) provides an intermediate level of
service between IP numbers and domain names. A handle adheres
permanently to an owner, who may assign and reassign it to different
addresses at will. But a handle is a number, carrying no significance
in natural language. Any user desiring a handle may generate one from
a public key. This memo describes a simple implementation of an Open
Network Handle System using the security extensions to the Domain Name
System (DNSSEC).

\end{abstract}

\tableofcontents

\section{Introduction}

Open Network Handle System (ONHS) Handles are hierarchical lists of
tokens, much like domain names except that individual labels are
numerical, and carry no significance in natural language. Certain
handle labels are constrained to contain certain types of
cryptographic public keys.

\subsection{Handles Provide Persistence, Not Meaning}

A conventional domain name provides two different types of value:
\begin{itemize}
\item It provides persistent reference to a particular network agent as the
agent's IP address changes.
\item It has some human meaning related to the owner, making it easier than
a meaningless token to remember, communicate, and guess.
\end{itemize}

Handles are intended to provide only the first type of value:
persistent reference. Because they carry no intuitive human meaning,
all handles are essentially equally valuable. Handle ownership should
attract very little dispute. In the absence of dispute handle
assignment may be completely automated, and human administration of
the handle system may be minimized. Most handle owners will probably
take steps outside of ONHS to connect their handles to domain names
and/or other sources of human meaning.

The proposed Open Network Handle System is intended to provide the
minimal service needed so that individual users of the Internet may
enjoy the value of persistent reference through handles. Each handle
owner is entirely responsible for using public-key cryptography to
generate and defend her handles, and for announcing correct
information to resolve handles to addresses. ONHS only tries to
provide prompt and correct resolutions of handles to addresses with
high probability, in order to establish contact between parties
interested in communicating. All other issues must be addressed by the
parties themselves through their direct communications, or through
other services outside of ONHS.

In particular, ONHS does not certify the correctness of individual
resolutions. Queriers and handle owners are entirely responsible for
establishing whatever level of authenticity they require. Since ONHS
handle are hash codes of cryptographic public keys, queriers and
handle owners may choose to use that key for their own authentication
if they deem the threats to be relatively light. This separation of
responsibility allows each pair of corresponding querier and handle
owner to determine their own criteria for satisfactorily authentic
communication.

From the point of view of users, ONHS provides a location service, not
an authentication service. ONHS uses authentication techniques in its
operations, but they are there to improve the rate of correct
resolutions, not to guarantee authenticity of individual resolutions.

In the presence of an ONHS system that resolves handles to addresses,
and other systems, such as the Domain Name System (DNS), that resolve
meaningful names, we may bind names to handles instead of to
addresses. The name systems may concentrate on issues to do with
meaning at any given time. The persistence of handles provides no
meaning on its own, but allows users to accumulate meaning over long
sequences of transactions, including transactions involving the
resolution of meaningful names into handles.

\subsection{Handle Values}

A handle value encodes a complete method for authenticating bindings
to the handle. Handle values are hierarchical sequences of handle
labels, analogous to domain name values. A useful implementation
should support at least two types of handle labels:
\begin{itemize}
\item A hashed public-key (PK) label consists of a code indicating that it
is a PK label, a code describing a public-key authentication algorithm
and a hash function, along with the hash code of the public key.
\item An inherited-authority (IA) label inherits its authentication from the
nearest ancestor of another type. It consists of a code indicating
that it is an IA label, along with an arbitrary bit string
distinguishing it from sibling labels.
\end{itemize}
To attract users who cannot yet manage cryptographic keys effectively,
we should offer a third type of handle.
\begin{itemize}
\item An inherited out-of-band authenticated (OA) label is authenticated by
a third-party who checks the authenticity of updates from a putative
owner outside of ONHS, probably through some sort of password
scheme. It consists of a code indicating that it is a OA label, along
with an arbitrary bit string distinguishing it from sibling
labels. The nearest ancestor with a type other than IA determines the
authentication method of the trusted third-party.
\end{itemize}
The authentication method associated with a handle is the one
specified by the lowest label in the hierarchy.

\subsection{Handle Operations}

\subsubsection{Updates on Handles}

The owner of a handle is the agent who is able to authenticate
updates, typically by knowing the secret key corresponding to the
public key whose hash code is embedded in the handle value. A handle
owner may perform the following operations:
\begin{itemize}
\item create a new handle;
\item assign an address temporarily to a handle;
\item delegate a handle temporarily to another handle, possibly with a
different owner;
\item cancel a handle irrevocably;
\item transfer a handle irrevocably to another handle, usually with a
different authentication key;
\item mark a handle's security irrevocably as compromised.
\end{itemize}
Each handle assignment, delegation, and transfer must be authenticated
according to the authentication method associated with that
handle. Each handle creation must be authenticated according to the
authentication method associated with its parent. It makes sense to
have two different sorts of cancellations, one authenticated by the
handle and the other by its parent.

Within a contiguous zone of handles all but the root having
inherited-authentication, handle and parent authorities are the
same. The handle/parent authority distinction is only important at the
boundaries of zones of authentication authority.

\subsubsection{Resolution Queries on Handles}

Any participant in the network may query any handle to determine the
address that it resolves to. Resolution should follow delegations and
transfers on ancestors of the handle that is queried, as well as
delegations and transfers at leaf handles, until it reaches an
address. Delegations and transfers work just like DNAME delegations in
DNS.

\subsubsection{Auditing Operations on Handles}

To guard against errors and misbehavior by name servers, and against
evildoers spoofing name servers and/or handle owners, ONHS should be
as publicly auditable as feasible. On an item-by-item basis, any party
may query a name server for an individual record of creation,
assignment, delegation, cancellation, transfer, or compromise
announcement, along with its owner-provided certificate of
authenticity. Any party may also request that a handle server
authenticate a message. But authentication by a handle server only
gives assurance that the message indeed came from that server, not
that the message is correct. Such queries may be used for spot
checking, and also for retracing all of the steps in a suspect
resolution of a handle to an address.

Interested parties may request complete dynamic audit trails for
particular handles. To provide such an audit trail, a handle server
should forward all authentic updates and inauthentic attempted updates
to the given handle as it receives them. If an onerously large number
of parties request dynamic audit trails, at least one should be
granted, with priority to the trail if any requested by the handle
owner.

To the extent it is feasible and affordable, handle servers should
keep archival logs of transactions for retrospective audits.

Audit operations are intended to support the integrity of handle
resolution. The ONHS is responsible for resolving handles promptly to
correct addresses with high probability, and with keeping the rate of
erroneous resolutions low. ONHS performance depends on a reasonable
level of enforcement of authenticity, since inauthentic updates to
handles produce erroneous resolutions and prevent correct ones. But
ONHS performance does not require, nor is it intended to provide, a
high assurance of authenticity for an individual resolution of a
handle to an address.

Queriers and handle owners should take their own end-to-end steps to
achieve a satisfactory level of assurance of quality for each of their
mutual transactions. When queriers and handle owners judge their
vulnerability to be low, they may use public keys taken from handles
for these purposes. But when assurance of authenticity and/or other
qualities is important, and when there is a serious threat of attack,
they should use independently stored additional cryptographic keys,
and other resources that they obtain outside of ONHS. In particular,
correspondents who engage in a series of important communications
should use ONHS only to make efficient connections. They should store
additional cryptographic keys and other resources to authenticate
their communications completely independently.

\section{Recommended Uses of Handles}

\subsection{Organizing Handles to Reflect Authority Hierarchies}

While domain name hierarchies reflect both hierarchies of authority
and hierarchies of meaning, handle hierarchies should reflect only
hierarchies of authority. Therefore a typical handle owner should own
one top-level public-key handle for each public key that she wishes to
use, plus one level of inherited-authority handles below each
public-key handle. Figure 1 shows the recommended two-level handle
hierarchy. PK 1 and PK 2 are public-key handles, while IA 11, IA 12,
IA 1m, IA 21, IA 22, IA 2n are inherited-authority handles.

\begin{figure}
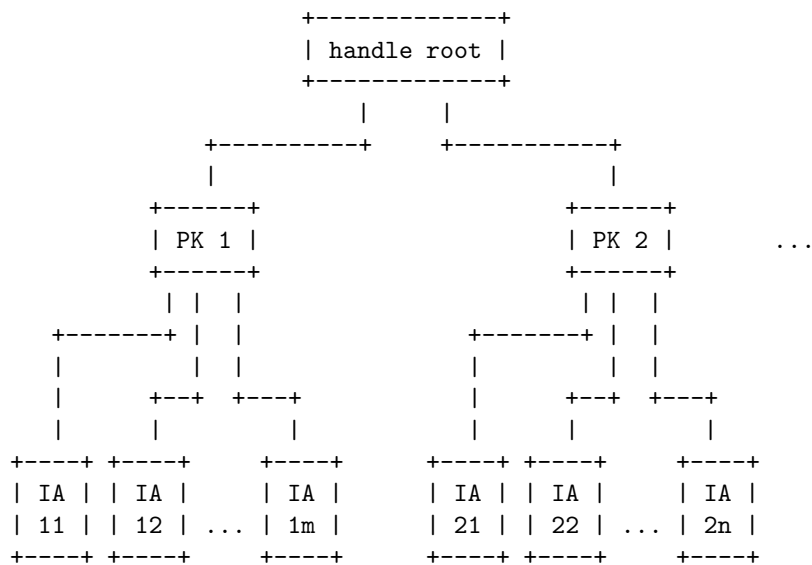

\begin{verbatim}
                     +-------------+
                     | handle root |
                     +-------------+
                         |     |
              +----------+     +-----------+
              |                            |
          +------+                      +------+
          | PK 1 |                      | PK 2 |       ...
          +------+                      +------+
           | |  |                        | |  |
   +-------+ |  |                +-------+ |  |
   |         |  |                |         |  |
   |      +--+  +---+            |      +--+  +---+
   |      |         |            |      |         |
+----+ +----+     +----+      +----+ +----+     +----+
| IA | | IA |     | IA |      | IA | | IA |     | IA |
| 11 | | 12 | ... | 1m |      | 21 | | 22 | ... | 2n |
+----+ +----+     +----+      +----+ +----+     +----+
\end{verbatim}
\caption{Recommended 2-level handle hierarchy}
\end{figure}

\subsubsection{Delegation of Authority}

When a handle owner wishes to delegate signing authority for a
subspace of handles, it is usually better to delegate one of her
second-level IA handles to another agent's second-level IA handle. A
handle owner may allow another public-key handle as a descendant of
her own public-key handle, unmediated by delegation, but this allows
less flexibility and I do not expect it to be common. The target of
delegation should introduce a third level of inherited-authority
handles. Figure 2 shows such a delegation as a horizontal barred arrow
(\texttt{==>}).

\begin{figure}
\begin{verbatim}
                     +-------------+
                     | handle root |
                     +-------------+
                         |     |
              +----------+     +-----------+
              |                            |
          +------+                      +------+
          | PK 1 |                      | PK 2 |       ...
          +------+                      +------+
           | |  |                        | |  |
   +-------+ |  |                +-------+ |  |
   |         |  |                |         |  |
   |      +--+  +---+            |      +--+  +---+
   |      |         |            |      |         |
+----+ +----+     +----+      +----+ +----+     +----+
| IA | | IA |     | IA |=====>| IA | | IA |     | IA |
| 11 | | 12 | ... | 1m | del  | 21 | | 22 | ... | 2n |
+----+ +----+     +----+      +----+ +----+     +----+
                                | |
                                | +---------+
                                |           |
                             +-----+     +-----+
                             | IA  |     | IA  |
                             | 211 | ... | 21p |
                             +-----+     +-----+
\end{verbatim}
\caption{Recommended delegation of authority}
\end{figure}

Figure 3 shows the hierarchy of handles under PK 1 seen by a naive
querier who only resolves handles to addresses, ignoring
authority. But bindings for IA 211 through IA 21p are authorized by
the owner of PK 2, while those for IA 11, IA 12, ... IA 1m are
authorized by the owner of PK 1.

\begin{figure}
\begin{verbatim}
                  |
              +------+
              | PK 1 | ...
              +------+
               | |  |
       +-------+ |  |
       |         |  |
       |      +--+  +---+
       |      |         |
    +----+ +----+     +----+
    | IA | | IA |     | IA |
    | 11 | | 12 | ... | 1m |
    +----+ +----+     +----+
                       | |
                       | +---------+
                       |           |
                       +-----+     +-----+
                       | IA  |     | IA  |
                       | 211 | ... | 21p |
                       +-----+     +-----+
\end{verbatim}
\caption{Visible handle hierarchy from Figure 2}
\end{figure}

Further delegations can create an arbitrarily deep visible handle
hierarchy with only three levels of hierarchy in the delegationless
structure. Notice that depth in such a visible hierarchy comes only
from the hierarchy of delegation of authority. Other organizational
hierarchy may be indicated outside of the handle system.

Three levels of delegationless structure, along with delegations from
the second and third levels to the second level, suffice to represent
an arbitrary tree-shaped hierarchy of authority. The sources of
delegations may be arbitrarily deep in the visible hierarchy, but the
target of a delegation should always be a newly created handle at the
second level of the delegationless structure.

Three delegationless levels, plus delegations similar to those in
Figures 1-4, suffice for all of the authority structures that occur to
me so far. Figure 4 shows a structure in which the owner of PK 2
exercises some authority delegated by PK 1, some authority delegated
by PK 3, and some authority that simultaneously serves PK 1 and PK
3. IA 212 and IA 231 both delegate to IA 22. There should be more
handles below IA 22, but I ran out of space in the picture.

\begin{figure}
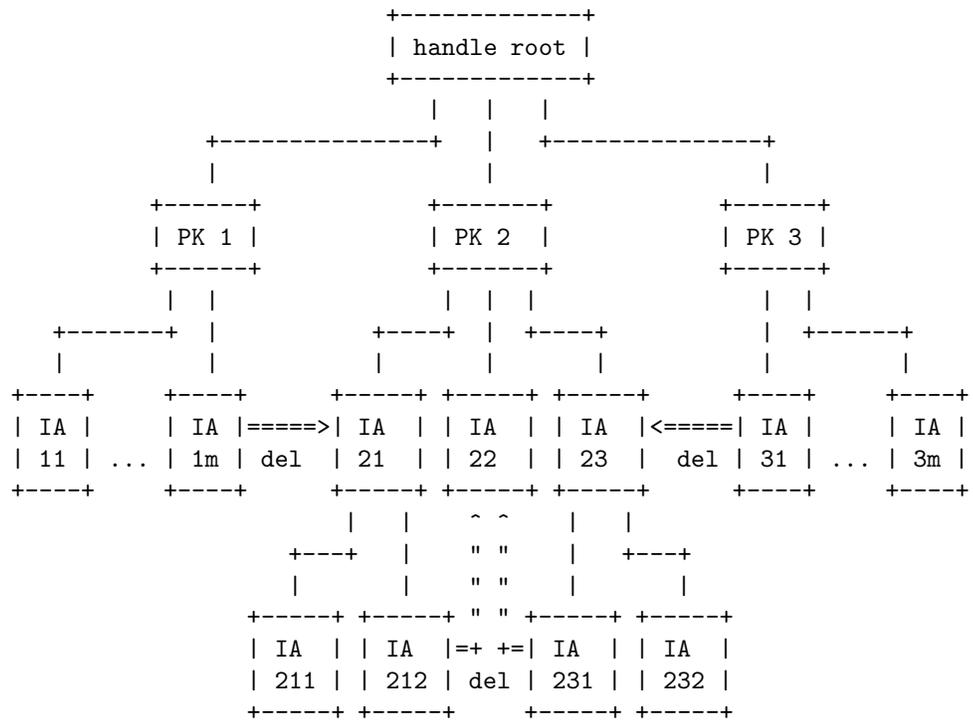

\begin{verbatim}
                           +-------------+
                           | handle root |
                           +-------------+
                              |   |   |
              +---------------+   |   +---------------+
              |                   |                   |
          +------+            +-------+            +------+
          | PK 1 |            | PK 2  |            | PK 3 |
          +------+            +-------+            +------+
           |  |                |  |  |                |  |
   +-------+  |           +----+  |  +----+           |  +------+
   |          |           |       |       |           |         |
+----+     +----+      +-----+ +-----+ +-----+      +----+     +----+
| IA |     | IA |=====>| IA  | | IA  | | IA  |<=====| IA |     | IA |
| 11 | ... | 1m | del  | 21  | | 22  | | 23  |  del | 31 | ... | 3m |
+----+     +----+      +-----+ +-----+ +-----+      +----+     +----+
                        |   |    ^ ^    |   |
                    +---+   |    " "    |   +---+
                    |       |    " "    |       |
                 +-----+ +-----+ " " +-----+ +-----+
                 | IA  | | IA  |=+ +=| IA  | | IA  |
                 | 211 | | 212 | del | 231 | | 232 |
                 +-----+ +-----+     +-----+ +-----+
\end{verbatim}
\caption{Overlapping delegations of authority}
\end{figure}

I recommend that users go as far as possible with the minimum number
of levels: two for many purposes, three for delegations of
authority. But it seems prudent to leave the system open to more
levels for purposes conceived in the future.

\subsubsection{Password Authentication Using an OA Hierarchy}

As soon as public-key software with easy user interfaces is widely
deployed, we should abandon the use of password authenticated
handles. But for a user who would like the benefit of a handle, and
whose handle is not valuable enough to others to invite attack,
password authentication is a good interim method. The implementor of a
password authentication hierarchy may use email confirmation or other
techniques to improve the security of password-authenticated updates.

An agent who wishes to offer password authenticated handles should
claim a new public-key handle and create a two-level hierarchy of the
PK handle with OA handles as children, as shown in Figure 5.

\begin{figure}
\begin{verbatim}
                       +-------------+
                       | handle root |
                       +-------------+
                              |
                              |
                              |
                          +------+
                          | PK 1 | ...
                          +------+
                           | |  |
                   +-------+ |  |
                   |         |  |
                   |      +--+  +---+
                   |      |         |
                +----+ +----+     +----+
                | OA | | OA |     | OA |
                | 11 | | 12 | ... | 1m |
                +----+ +----+     +----+
                 |  |
             +---+  +----+
             |           |
          +-----+     +-----+
          | IA  |     | IA  |
          | 211 | ... | 21p |
          +-----+     +-----+
\end{verbatim}
\caption{Recommended password-authentication hierarchy}
\end{figure}

ONHS treats OA handles exactly the same as IA handles, since it only
has access to public-key signatures and not to the password file. But
an agent who operates a password-authenticated handle service, and who
takes no responsibility for the behavior of the OA handle owners,
should mark those handles OA to warn the public that her PK authority
is only intended to authenticate the receipt of updates with correct
passwords, not to take credit or blame for the behavior of the agent
with the OA handle.

Each OA handle owner should establish a level of IA handles, just like
a PK handle owner.

A corporate agent may use password authentication internally to
delegate authority for updates on certain handles to subordinates. As
long as the corporate agent wants credit for the behavior of the
subordinates, and accepts the risk of blame, she should mark the
subordinate handles as IA. The IA vs.\ OA distinction indicates the
authority relationship, rather than the technical mechanisms used in
generating authentic updates.

A querier who discovers an address for an OA handle should not use the
public key associated with the signature on handle updates for
communications with the handle owner. Such a use is reasonable for PK
and IA handles.

\subsubsection{Upgrading a Key or Selling a Handle by Irrevocable Transfer}

Although irrevocable transfer of a handle is essentially the same as
temporary delegation in terms of the data structure that represents
it, the natural use of irrevocable transfer is quite different, and
it calls for different policy.

\paragraph{2.1.3.1 Upgrading a Key\\}

The technical device of using a public key as a handle works only as
long as the key, and the cryptographic technique that it uses, remain
reasonably secure. Over a long period of time, a handle owner will
often need to create one long-lived virtual handle by connecting old
handles to newer ones as the old ones become obsolete. For handles
with relatively low traffic and low commercial value, individual keys
may be viable longer than common security practice suggests. But it is
likely that all cryptographic techniques (at given key sizes) known
today will eventually become obsolete due to advances in mathematics
and increases in computing speed.

There appears to be no perfect method for transferring authority from
an obsolescent key to a new and stronger one. The ONHS can provide key
owners with a significant window of time for such a transfer, and can
help to advertise the transfer to queriers.

Long before a given key K1 becomes compromised or otherwise obsolete, and
as soon as a stronger encryption technique (or just a longer key)
becomes practicable, the owner of K1 should perform the following
steps to transfer it.

\begin{itemize}
\item Claim a new public-key handle with key K2, using the strongest encryption that is
affordable at the time.
\item Copy the hierarchy of inherited-authority handles below K1 to a
corresponding hierarchy below K2. Keep the numerical values for the
copied handles the same, not just the topological structure.
\item Replicate the address bindings, delegation structure, and any other
information associated with the hierarchy below K1 in the hierarchy
below K2.
\item Redirect delegations from the same owner's other handles to use the
hierarchy below K2 in place of the hierarchy below K1. In principle,
this step may be delayed as long as K1 remains usable, but it seems
best to do it now.
\item Contact other handle owners who have delegated authority into the
hierarchy below K1, and instruct them to redirect to the hierarchy
below K2. The owner of K1/K2 may not know the identities of all such
other owners, but she probably knows those who are most important to
her. In principle this step may be delayed, or even left to the
diligence of the other handle owners, but it is probably best in most
cases to do it now.
\item Verify through test queries that the hierarchy below K2 is
satisfactory.
\item Perform an irrevocable transfer of K1 to K2.
\end{itemize}

Because the impact of irrevocable transfer on pure handle-to-address
resolution is the same as the impact of temporary delegation, queriers
of the hierarchy below K1 will be forwarded to the hierarchy below
K2. But handle servers should also report the transfer to each
querier, and each querier should replace all current references to K1
with references to K2.

The owner of K1/K2 should also advertise the transfer through any
other available channels and encourage all correspondents to update
references from K1 to K2. She may use an audit trail of queries on K1
to discover naive queriers and encourage them to update to K2. If ONHS
gains widespread use, application software (such as Web browsers)
should automate update of transferred handles, probably with
notification to and confirmation by the user.

Later, when the owner of K1/K2 decides that the appropriate balance
between risk of compromise to K1 and risk of losing correspondents has
passed, she should:
\begin{itemize}
\item Cancel K1 irrevocably.
\end{itemize}
If she believes that K1 has been compromised, then she should also
mark it as compromised.

Notice that temporary delegation should normally be applied to an IA
handle below a PK handle, but irrevocable transfer for key upgrade
should normally be applied to a PK handle. Figure 6 shows the
recommended structure for key upgrade.

\begin{figure}
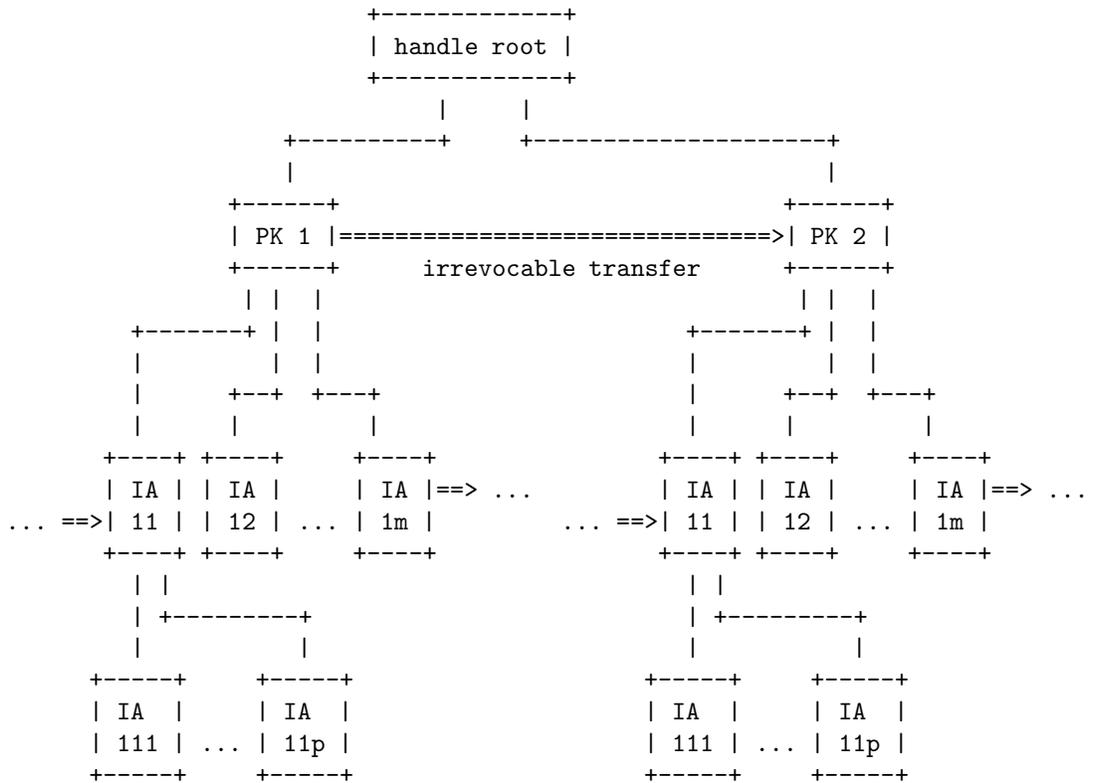

\begin{verbatim}
                           +-------------+
                           | handle root |
                           +-------------+
                                |     |
                     +----------+     +---------------------+
                     |                                      |
                 +------+                                +------+
                 | PK 1 |===============================>| PK 2 |
                 +------+      irrevocable transfer      +------+
                  | |  |                                  | |  |
          +-------+ |  |                          +-------+ |  |
          |         |  |                          |         |  |
          |      +--+  +---+                      |      +--+  +---+
          |      |         |                      |      |         |
        +----+ +----+     +----+                +----+ +----+     +----+
        | IA | | IA |     | IA |==> ...         | IA | | IA |     | IA |==> ...
 ... ==>| 11 | | 12 | ... | 1m |         ... ==>| 11 | | 12 | ... | 1m |
        +----+ +----+     +----+                +----+ +----+     +----+
          | |                                     | |
          | +---------+                           | +---------+
          |           |                           |           |
       +-----+     +-----+                     +-----+     +-----+
       | IA  |     | IA  |                     | IA  |     | IA  |
       | 111 | ... | 11p |                     | 111 | ... | 11p |
       +-----+     +-----+                     +-----+     +-----+
\end{verbatim}
\caption{Recommended transfer for key upgrade}
\end{figure}

\paragraph{2.1.3.2 Selling a Handle\\}

Irrevocable transfer may also be used to transfer a handle to a new
owner, for example as part of the sale of a portion of a company's
business. In this case, the steps outlined in 2.1.3.1 above divide
naturally into those performed by the donor and those performed by the
recipient. In the case of the sale of a portion of a company's
business, the transfer will usually go from a subsidiary IA handle in
the donor's hierarchy to a subsidiary IA handle in the recipient's
hierarchy. The recipient should verify by a test query that the donor
actually executed the transfer. Since the recipient has an incentive
to enforce the transfer, while the donor may have no such incentive,
or even a counterincentive, the recipient should archive a signed
record of the transfer, and should audit activity on the donor
key. For transfers of valuable handles, the parties should use
certification services outside of ONHS to record and enforce
conditions of the transfer.

\subsection{Resolving Domain Names to Handles}

Hierarchies of domain names should reflect a variety of hierarchies of
meaning. Some hierarchies of meaning, such as an organizational
hierarchy for a corporation, may correspond closely with hierarchies
of authority. But many or most hierarchies of meaning, such as the
hierarchy of products and services offered to a customer, may have a
very different structure from the authority structure. Even structures
of corporate authority may not match up perfectly with the technical
authority over cryptographic keys for handles that drives the handle
hierarchy and delegation structure. So a user of both handles and
names should develop the structure of each independently, then provide
information to resolve names to handles. In a rough analogy to
programming language compilers, we may think of domain names as
analogous to identifiers or variable names, handles as analogous to
symbol table entries, and IP numbers as analogous to memory addresses.

When a user of both DNS and ONHS creates a new leaf domain name in
DNS, she should usually create a new corresponding handle to track its
value. That handle should usually not be a public-key handle, but
rather an inherited-authority handle below a public-key handle,
so that the holder of that public key may perform other roles
associated with other handles. Occasionally, there may already be a
handle appropriate to the new name. But users should make multiple
references from different names to the same handle only when the
different names are intentionally synonymous, not just accidentally
equivalent for the moment. For example, president.example.com and
treasurer.example.com should map to different handles representing
the different roles of the offices, even though a single person holds
both offices at the moment.

In Section 3 we see that a handle system may be implemented as a
contiguous set of zones in DNS. In that case the mapping of domain
names to handles may be accomplished with DNAME resource records.

On the other hand, the assignment of a handle to a name should change
very seldom. Changes in the way that the meaning of the name is
deployed in the world (a new person taking over the role referred to
by the name, the office referred to by the name moving to a new
building, ...) should usually be reflected by reassignments of the
handle. When example chooses a new president, the handle associated
with president.example.com should be redelegated or reassigned, rather
than the name president.example.com. The authority for that
redelegation presumably comes from example's board of directors.

The handle associated with a name should change only when there is a
change of cryptographic key, a permanent change of authority over the
handle, an essential change in the meaning of the name (e.g., a change
in the dictionary meaning of the name), or for some other reason the
structure of handle space must change. The ONHS is designed to allow
users to set up structures that avoid the need for restructuring
(address assignment and delegation don't count as restructuring), but
surely the system and the users will fail in some cases.

Although a change of cryptographic key (either because the old key is
overused, or to upgrade to stronger cryptographic techniques) and a
permanent change in the authority over a handle probably require a
change in the name-handle assignment, even that change is partly
supported by ONHS. In such a case, the owner of the handle should
announce a permanent transfer of the handle. Such a transfer affects
all future resolutions of the old handle to an address, but everyone
who refers to the old handle should also update that reference to use
the new handle as soon as he discovers the transfer. By contrast, a
temporary delegation of authority is completely transparent (except
for auditing purposes), and someone who queries a handle and discovers
a temporary delegation to another handle should usually not update his
own reference.

I re-emphasize that ONHS only provides a tool by which agents may
point to addresses by binding their handles, and queriers may locate
those addresses by querying handle servers to resolve those
handles. ONHS does nothing to ensure the semantic correctness of the
resolvent. If example.com binds president.example.com to a handle that
is intended to track the company's president, ONHS does nothing to
help make sure that the handle is indeed bound to the legitimate
president. It merely resolves the handle according to whatever
correctly signed bindings it has received.

Notice that delegations and transfers in ONHS are exposed to the same
potential for circularity as DNAME delegations in DNS. But a policy of
always delegating to a newly created IA child avoids cycles among
handles, even when there is a cycle of delegation among handle
owners. Figure 7 shows a cyclic delegation of authority between the
owners of PK 1 and PK 2, with no actual cycle of delegation among
handles.

\begin{figure}
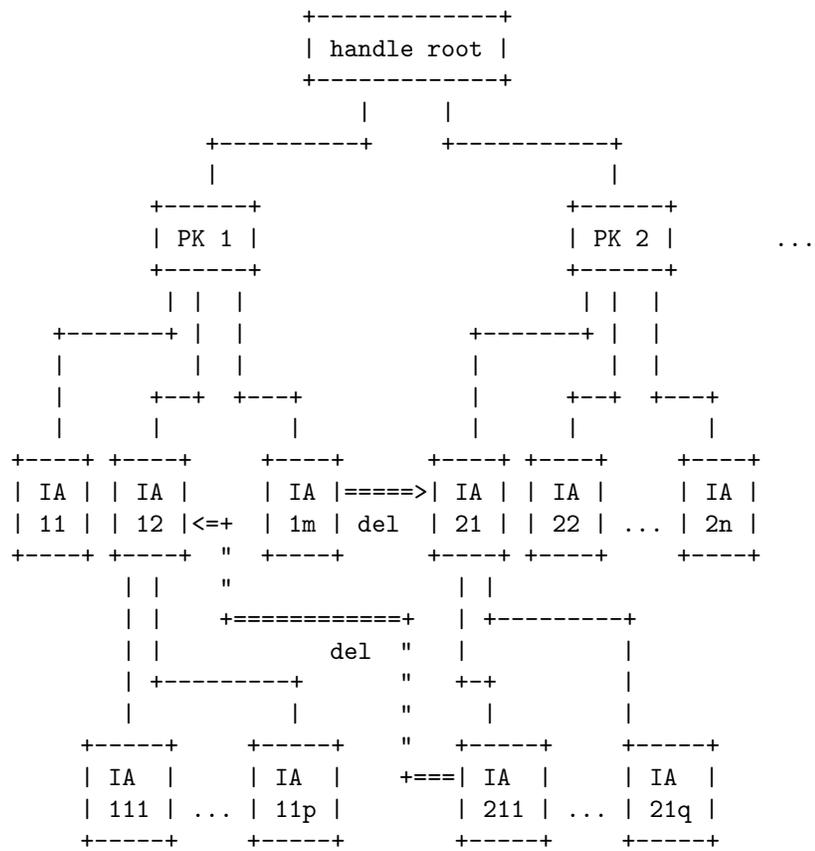

\begin{verbatim}
                     +-------------+
                     | handle root |
                     +-------------+
                         |     |
              +----------+     +-----------+
              |                            |
          +------+                      +------+
          | PK 1 |                      | PK 2 |       ...
          +------+                      +------+
           | |  |                        | |  |
   +-------+ |  |                +-------+ |  |
   |         |  |                |         |  |
   |      +--+  +---+            |      +--+  +---+
   |      |         |            |      |         |
+----+ +----+     +----+      +----+ +----+     +----+
| IA | | IA |     | IA |=====>| IA | | IA |     | IA |
| 11 | | 12 |<=+  | 1m | del  | 21 | | 22 | ... | 2n |
+----+ +----+  "  +----+      +----+ +----+     +----+
        | |    "                | |
        | |    +============+   | +---------+
        | |            del  "   |           |
        | +---------+       "   +-+         |
        |           |       "     |         |
     +-----+     +-----+    "   +-----+     +-----+
     | IA  |     | IA  |    +===| IA  |     | IA  |
     | 111 | ... | 11p |        | 211 | ... | 21q |
     +-----+     +-----+        +-----+     +-----+
\end{verbatim}
\caption{Authority circularity without handle circularity}
\end{figure}

\section{Handle Zones in DNS}

A name server in DNS may act as an ONHS handle server essentially by
restricting its behavior on one or more zones to support only valid
handle operations. A "handle zone" is any DNS zone in which domains
are restricted to be handles. The technical requirements of ONHS match
up very closely with the standards for DNS [DNS] with the security
extensions [DNSSEC]. I noticed two points in which the technical
concepts of ONHS are skewed slightly (and I believe harmlessly) with
respect to the nearly analogous DNS concepts.

\begin{itemize}
\item DNS zones are simultaneously zones of authority over a portion of the
name space, and zones of responsibility for providing resolution
information within the same portion. In ONHS, a zone of contiguous
inherited-authority handles, rooted at a public-key handle, forms a
zone of authority for authenticating update transactions. It is
generally good management practice to keep responsibility and
authority in close correspondence, but resource availability will
probably lead to informal agreements and contracts whereby a
separately controlled handle server resolves all or part of a zone of
public-key authentication authority on behalf of the authority. In
such a case, the DNS zone corresponds to the zone of responsibility
for resolution, and the operator of the name server for the zone
merely enters all properly certified updates from the zone authority
in its tables. Proper treatment of irrevocable transfer requires that
the authentication authority zone and the resolution responsibility
zones diverge.
\item The DNS security extensions [DNSSEC] conceive of signatures as
certificates of authenticity provided by authoritative name
servers. ONHS conceives of its certificates as coming from handle
owners, who may not all be capable of operating their own handle
servers. But the nature of public-key certificates allows them to be
copied freely, so this is a difference in description and explanation
rather than in technical constraints. A DNS name server may store
certificates sent to it by other parties, after verifying their
authenticity with a public key.
\end{itemize}

There is also a radical difference between the underlying security
models motivating the design of DNSSEC vs.\ ONHS. DNSSEC is intended to
provide chains of trust. ONHS is intended to provide a completely
distributed security structure, with each handle owner an independent
security root. (See Section 6.1 below for a more detailed comparison.)
But the primitive operations on keys and signatures in DNSSEC are
flexible enough to support either model.

So, an ONHS root may essentially be implemented in a DNS zone by any
security-aware name server that is willing to administer the handle
root zone for names representing handles only. Handle owners may
establish zones immediately below the handle root zone. The handle
root server may limit himself to the top level of the handle zones,
and leave it to each handle claimant to provide a name server for the
inherited-authenticity zone beneath her public-key handle. Or, a
particularly altruistic root server may also maintain separate zone
files for some or all handle claimants,

\subsection{An Altruistic Handle Root Server}

Suppose that an altruistic sponsor is willing to support a public
handle system at \texttt{handleroot.example.org}. The example sponsor is
willing to provide two security-aware name servers [DNSSEC] for the
handle root zone containing public-key handles, and possibly for zones
of inherited-authenticity handles below the public-key handles of
those handle owners who request it. It would be a nice piece of PR to
run a handle root server at a top-level domain in DNS, such as \texttt{z.}\ or
\texttt{handles.}\ or \texttt{handleroot}. A second-level domain such as \texttt{handleroot.org}.
would be slightly less nice. But any stable domain that doesn't
consume too many characters in the maximum length of domain names will
work.

The altruistic sponsor fulfills all correctly signed requests
uncritically. It does not check, nor take any responsibility for, the
identity or good behavior of handle owners. Handle owners may register
with other systems, or use direct communications with their
correspondents, to establish identity or other qualities.

A very altruistic sponsor, who might be the same as the handleroot
sponsor, or different, may claim a top-level public key handle and
implement a cheap or free public password-authenticated handle service
below it, marking the password-authenticated handles as OA
(out-of-band authentication). Such a sponsor should use well known
techniques for receiving password-authenticated transactions through
Web forms or by email, and perhaps confirming them further by email or
other channel. Such a service may be crucial to early dissemination of
ONHS use. But it makes no special technical demands on a DNS
implementation, so I do not treat it further in this article.

A less altruistic agency may run a proprietary handle server, with its
own handle root, by supporting whatever subset of the altruistic
operations it chooses for whatever constituency it likes.

\subsection{Representing Handles by Domain Names}

Each public-key handle is represented by a domain name label of the form

\begin{verbatim}
h1g<mmm>k<n...n>
\end{verbatim}

\begin{itemize}
\item The initial '\texttt{h}' indicates that this is a handle, and satisfies the DNS
requirement that a label starts with an alphabetic character.
\item The '\texttt{1}' indicates that this is a public-key handle.
\item The '\texttt{g}' stands for 'algorithm'. ('\texttt{a}' might be confused with the
hexadecimal digit for 11.) 
\item \texttt{<mmm>} is a decimal number of 1 to 3 digits denoting the key algorithm
(with no leading zeroes), using the same numbering in \texttt{KEY} RRs
[DNSSEC].
\item The '\texttt{k}' stands for 'key'.
\item \texttt{<n...n>} is a suffix of the 40-digit SHA1 hash of the RSA public key in
hexadecimal (with leading zeroes, if any), containing at least 14
hexadecimal digits (at least 20 is strongly recommended, and there's
nothing wrong with using all 40).
\end{itemize}

\begin{example}
\textup{\texttt{h1g5k0061A38F9A3540B9.handleroot.example.org.}\ is
a public-key handle for the signature algorithm RSA/SHA1
[RSA/SHA1], supposing that the random number \texttt{0061A38F9A3540B9} is
actually the last 16 digits of the 40-digit hexadecimal notation for
the SHA1 [SHA1] hash of an RSA [RSA] public key.}
\end{example}

While handle syntax is unlikely to be used spontaneously for another
purpose, it is not intended to mark a domain name label unambiguously
as a handle. It is the position below \texttt{handleroot.example.org}, and
the general knowledge that this is a root domain for an implementation
of ONHS, that determines the intention to use the domain name as a
handle.

An inherited-authority handle is represented by a domain name label of
the form

\begin{verbatim}
h0k<n...n>
\end{verbatim}

\begin{itemize}
\item The initial '\texttt{h}' indicates that this is a handle, and satisfies the DNS
requirement that a label starts with an alphabetic character.
\item The '\texttt{0}' indicates that this is an inherited-authority handle.
\item The '\texttt{k}' stands for 'key'.
\item \texttt{<n...n>} is a decimal number of 1 to 60 digits without leading
zeroes.
\end{itemize}

\begin{example}
\textup{\texttt{h0k2.h0k3.h1g5k0061A38F9A3540B9.handleroot.example.org}\
  is
inherited-authority handle number 2 below inherited-authority
handle number 3 below the public-key RSA/SHA1 handle with key suffix
\texttt{0061A38F9A3540B9}.}
\end{example}

\subsection{Resource Records (RR) for Handle Bindings}

I demonstrate the implementation of ONHS in DNS resource records
through a paradigmatic example that appears to cover the essential
cases.

\subsubsection{RRs in the Handle Root Zone}

The handle root zone contains the following RRs, even before any
handles are claimed.

\begin{verbatim}
handleroot.example.org. IN SOA handleserver1 handlemaster (
   1 1d 1h 1w 1h
   )
handleroot.example.org. SIG SOA (
   5 3 86400 20050415223412 20050401223412
   handleroot.example.org.
   <sig>
   )

handleroot.example.org. IN NS handleserver1
handleroot.example.org. IN NS handleserver2
handleroot.example.org. SIG NS (
   <params>
    handleroot.example.org.
   <sig>
   )

handleserver1 IN A 183.021.254.010
handleserver1 SIG A (
   <params>
   handleroot.example.org.
   <sig>
   )

handleserver2 IN A 183.021.254.020
handleserver2 SIG A (
   <params>
   handleroot.example.org.
   <sig>
   )

handleroot.example.org. IN KEY 256 3 5 (
   AQPZrT453eyUm/kO7rOc0GIUd7PX3n2gueMFtIGOzSUOaOt4lmiJq7bo
   Fb2p9S2hXRyqZKoD82ouRxwRqEfApYyt
   )
\end{verbatim}

\begin{itemize}
\item \texttt{IN} indicates the Internet protocol. The alternatives are not very
relevant today.
\item The \texttt{SOA} (start of authority) record establishes a zone rooted at
\texttt{handleroot.example.org}.
\begin{itemize}
\item The name '\texttt{handleserver1}' indicates that
\texttt{handleserver1.handleroot.example.org} is the primary authoritative
handle/name server for the handle root zone.
\item The name '\texttt{handlemaster}' indicates that
\texttt{handlemaster@handleroot.example.org} is the email address of a
person who will deal with correspondence regarding the zone and its
name servers.
\item The '\texttt{1}' is a serial number that we increment whenever the zone data
change.
\item The '\texttt{1d}' specifies a modest period of one day for refreshing data at
slave servers (mirrors). This is a public service zone, and handle
owners who want real quick dissemination should run their own handle
servers.
\item The '\texttt{1h}' specifies a modest timeout of one hour to wait before
retrying if the server fails to respond.
\item The '\texttt{1w}' is a modest timeout of one week to wait before invalidating a
slave server that cannot reach an authoritative server.
\item The final '\texttt{1h}' is a default time-to-live of one hour for messages
indicating that a given handle cannot be resolved.
\end{itemize}
\item The \texttt{SIG SOA} record signs the \texttt{SOA} record with the handle root
operator's key.
\begin{itemize}
\item The '\texttt{5}' is the number for RSA/SHA1 signatures.
\item The '\texttt{3}' is the number of labels in \texttt{handleroot.example.org}.
\item The '\texttt{86400}' is a plausible initial \texttt{TTL} (see [DNSSEC] for the technical
reason why this is included) which has nothing special to do with
implementing ONHS.
\item The '\texttt{20050408223412}' indicates that this signature will expire at 12
seconds past 22:34 on 8 April 2005.
\item The '\texttt{20050401223412}' indicates that this signature was created at 12
seconds past 22:34 on 1 April 2005. So the signature is valid for one
week.
\item It's just tedious to create plausible looking signatures, so I just
use '\texttt{<sig>}' to represent the signature.
\end{itemize}
\item The two NS (name server) records repeat the primary \texttt{handleserver1} name
server and add the alternate server \texttt{handleserver2}.
\item The \texttt{SIG NS} record signs both of the \texttt{NS} records together. From now on,
I just use '\texttt{<params>}' to represent sensible parameters for the
signature.
\item The two \texttt{A} (address) records provide IP numbers for the name
servers. Logically, these are crucial, while the names for the name
servers are redundant. But RR syntax seems to require this indirect
specification.
\item The public key isn't strictly necessary for a handle server. It is
important for handle owners to authenticate their transactions on
their handles, but their authority does not derive from the root
handle server. But if the sponsor can afford it, the root handle
server should provide this defense against spoofing.
\item Similarly, the \texttt{SIG} records, signed with the root handle server's key,
are not strictly necessary, but probably helpful.
\end{itemize}

For each top-level public-key handle, the handle root zone should
contain a key record. Using for example three different RSA/SHA1 keys
(don't bother to check the hashing, it isn't correct):

\begin{verbatim}
h1g5k0061A38F9A3540B9 IN KEY (
   AQOp6Lb7uQyR+4FBiTZivr2xBm5ZQYRkNbcrVHZe/S0XUBSRWyVuQdH4
   DuaNnzdi/bywVFSvFCbLNcL724ECyqRV
   )
h1g5k93C1C124A3760B88 IN KEY (
   <key>
   )
h1g5kEFA0A37BB4260D3C IN KEY (
   <key>
   )
\end{verbatim}

This key record probably should not be signed by the handle root zone,
because DNSSEC is likely to mistake that for a certification that the
key should be trusted. (This is a tricky point, and should be
reconsidered. Depending on the behavior of other DNS name servers,
there might be a small value in authenticating that the key was
received by the handle server through normal channels.)

The handle root zone should contain records referring to the name
servers for each handle owner's zone. The domain names ns1 and ns2
used below are irrelevant, but appear to be required by DNS RR
syntax. ns1 and ns2 do not have to be in the handle owners' domains,
but we suppose that many handle owners have modest resources and don't
control any other domains. Normal zones require at least two name
servers, and preferably more. But for a small handle owner, one name
server might be plenty, and all that the owner's resources allow. I
show one and two name servers in the owner's domain, and three
outside, for examples.

\begin{verbatim}
h1g5k0061A38F9A3540B9 IN NS ns1.h1g5k0061A38F9A3540B9
h1g5k0061A38F9A3540B9 IN NS ns2.h1g5k0061A38F9A3540B9
h1g5k0061A38F9A3540B9 SIG NS (
   <params>
   h1g5k0061A38F9A3540B9
   <sig>
   )

ns1.h1g5k0061A38F9A3540B9 IN A 192.253.254.21
h1g5k0061A38F9A3540B9 SIG A (
   <params>
   h1g5k0061A38F9A3540B9
   <sig>
   )

ns2.h1g5k0061A38F9A3540B9 IN A 192.253.254.22
h1g5k0061A38F9A3540B9 SIG A (
   <params>
   h1g5k0061A38F9A3540B9
   <sig>
   )

h1g5k93C1C124A3760B88 IN NS ns1.h1g5k93C1C124A3760B88
h1g5k93C1C124A3760B88 SIG NS (
   <params>
   h1g5k93C1C124A3760B88
   <sig>
   )

ns1.h1g5k93C1C124A3760B88 IN A <ip number>
h1g5k93C1C124A3760B88 SIG A (
   <params>
   h1g5k93C1C124A3760B88
   <sig>
   )

h1g5kEFA0A37BB4260D3C IN NS exampleserver1.example.com
h1g5kEFA0A37BB4260D3C IN NS exampleserver2.example.com
h1g5kEFA0A37BB4260D3C IN NS exampleserver3.example.com
h1g5kEFA0A37BB4260D3C SIG NS (
   <params>
   h1g5kEFA0A37BB4260D3C
   <sig>
   )

exampleserver1.example.com IN A <ip number>
h1g5kEFA0A37BB4260D3C SIG A (
   <params>
   h1g5kEFA0A37BB4260D3C
   <sig>
   )

exampleserver2.example.com IN A <ip number>
h1g5kEFA0A37BB4260D3C SIG A (
   <params>
   h1g5kEFA0A37BB4260D3C
   <sig>
   )

exampleserver3.example.com IN A <ip number>
h1g5kEFA0A37BB4260D3C SIG A (
   <params>
   h1g5kEFA0A37BB4260D3C
   <sig>
   )
\end{verbatim}
These records are signed by the handle owners. They are called "glue"
records.

\begin{itemize}
\item For relatively low-value and low-traffic handles, it may be reasonable
to assign validity periods much longer than usual in the \texttt{<params>}.
\end{itemize}

Almost all records binding handles to addresses and delegating handles
to other handles should be stored with the handle owner's zone, with
no direct support from the handle root zone. But irrevocable transfer,
cancellation, and announcement of compromise should be stored
permanently in the handle root zone as well, since the handle owner
cannot necessarily be trusted to maintain them, and she is not the
only agent risking harm if they are lost. The target of an irrevocable
transfer should also archive the signed DNAME record transferring the
handle, but to make the record effective, it must be entered in zones
that will be queried by those who still know the the source handle.

Here are records transferring the subhandle
\texttt{h0k1.h1g5k0061A38F9A3540B9}, canceling \texttt{h0k1.h0k2.h1g5k0061A38F9A3540B9}
presumably because it is no longer useful, and canceling
\texttt{h1g5k93C1C124A3760B88} with an announcement that it is compromised.

\begin{verbatim}
h0k1.h1g5k0061A38F9A3540B9 DNAME h0k427.h1g5kEFA0A37BB4260D3C
h0k1.h1g5k0061A38F9A3540B9 SIG DNAME (
   <params>
   h1g5k0061A38F9A3540B9
   )

h0k1.h0k2.h1g5k0061A38F9A3540B9 IN A <impossible address> 
h0k1.h0k2.h1g0061A38F9A3540B9 SIG A (
   <params>
   h1g5k0061A38F9A3540B9
   )

h1g5k93C1C124A3760B88 IN TXT "Compromised 01/04/2003"
h1g5k93C1C124A3760B88 SIG TXT (
   <params>
   h1g5k93C1C124A3760B88
   )

h1g5k93C1C124A3760B88 IN A <impossible address>
h1g5k93C1C124A3760B88 SIG A (
   <params>
   h1g5k93C1C124A3760B88
   )
\end{verbatim}

The handle root server should return a handle's transfer, cancel, and
compromise records in response to every query on that handle and its
descendants.

In principle, additional signatures should be added here, but DNSSEC
doesn't appear to allow them (see Section 3.6). The signatures on
irrevocable operations - transfers, cancellations and compromise
announcements - may have much longer periods of validity than is
usually recommended. They should be preserved even when they expire
(eventually in a less accessible archive), because they represent the
best available information, even though the original signer may be
unable or unwilling to sign them again.

To produce authenticated negative responses to resolution queries, the
handle root zone should contain a complete set of signed \texttt{NXT}
records. These are not as valuable as the \texttt{NXT} records in the handle
owner's zone, signed by the handle owner. They only authenticate the
fact that the root handle server is not aware of particular records.

\subsubsection{RRs in the Handle Owner's Zone}

Each handle owner's zone should contain its own \texttt{SOA}, \texttt{NS} and \texttt{A} records
for its own name servers. The \texttt{NS} and \texttt{A} records are essentially the
same as the corresponding glue records in the handle root zone. I show
only the example of the \texttt{h1g5k0061A38F9A3540B9} zone.

\begin{verbatim}
h1g5k0061A38F9A3540B9.handleroot.example.org. IN SOA ns1 hm (
    1 1h 1m 1d 1m
    )
h1g5k0061A38F9A3540B9.handleroot.example.org. SIG SOA  (
    <params>
    h1g5k0061A38F9A3540B9.handleroot.example.org.
    <sig>
    )

h1g5k0061A38F9A3540B9.handleroot.example.org IN NS ns1
h1g5k0061A38F9A3540B9.handlreoot.example.org IN NS ns2
h1g5k0061A38F9A3540B9 SIG NS (
   <params>
   h1g5k0061A38F9A3540B9.handleroot.example.org.
   <sig>
   )

ns1 IN A 192.253.254.21
h1g5k0061A38F9A3540B9 SIG A (
   <params>
   h1g5k0061A38F9A3540B9.handleroot.example.org.
   <sig>
   )

ns2 IN A 192.253.254.22
h1g5k0061A38F9A3540B9 SIG A (
   <params>
   h1g5k0061A38F9A3540B9.handleroot.example.org.
   <sig>
   )
\end{verbatim}

For each leaf of the inherited-authority zone below the public-key
handle, the handle owner's zone should contain an \texttt{A} record and a
corresponding \texttt{SIG} (signature), all signed by the handle owner's
key. This is entirely the responsibility of the handle owner. A handle
owner's mistakes affect the value of her own handles, but not the
integrity of the ONHS.

\begin{verbatim}
h0k2.h0k3.h1g5k0061A38F9A3540B9 IN A 192.253.254.63
h0k2.h0k3.h1g5k0061A38F9A3540B9 SIG A (
   <params>
   h1g5k0061A38F9A3540B9.handleroot.example.org.
   <sig>
   )

h0k3.h0k3.h1g5k0061A38F9A3540B9 IN A 192.253.254.65
h0k3.h0k3.h1g5k0061A38F9A3540B9 SIG A (
   <params>
   h1g5k0061A38F9A3540B9.handleroot.example.org.
   <sig>
   )
\end{verbatim}

\begin{itemize}
\item \texttt{192.253.254.63} \texttt{192.253.254.65} are fictional IP numbers that the handle
owner might assign to these handles.
\end{itemize}

A querier's confidence in a handle resolution does not derive from
trust in the root handle server, nor in the handle owner's server, but
from the correspondence between the handle and the key with which it
is signed. The querier must know that the domain name is intended as a
handle, and must know the rule of correspondence between handles and
keys, in order to authenticate the handle meaningfully.

For good netizenship, the handle owner's zone should include records
announcing irrevocable transfers, cancellations, and announcements of
compromised keys. But since lack of these records may harm others
besides the handle owner, they should also be stored in the handle
root zone. Here are examples of appropriate good netizen records for
the \texttt{h1g5k0061A38F9A3540B9} zone, duplicating the information in the
handle root zone above.

\begin{verbatim}
h0k1.h1g5k0061A38F9A3540B9 DNAME h0k427.h1g5kEFA0A37BB4260D3C
h0k1.h1g5k0061A38F9A3540B9 SIG DNAME (
   <params>
   h1g5k0061A38F9A3540B9
   )

h0k1.h0k2.h1g5k0061A38F9A3540B9 IN A <impossible address> 
h0k1.h0k2.h1g0061A38F9A3540B9 SIG A (
   <params>
   h1g5k0061A38F9A3540B9
   )
\end{verbatim}

To provide useful negative information, the handle owner should
provide complete signed \texttt{NXT} (next) records as well. The handle owner's
signed \texttt{NXT} records provide the authoritative indication that certain
handles do not exist.

\subsection{Operating the Root Handle Server with BIND}

Normal bind software, versions 9 and later with the security
extensions, can perform almost all update and query operations of the
root handle server. Handle owners may transmit appropriate signed RRs
to the root handle server, who stores them and removes outdated
records, all according to normal bind operations. But before accepting
a new \texttt{KEY} record for a PK handle, the root handle server must verify
that the handle value contains an appropriate suffix of the public
key. That operation is not supported by bind, but it is very simple to
add code for it. In principle, once the correct \texttt{KEY} record is
established, normal bind operations will maintain the use of that key
for operations on the PK handle and the IA hierarchy below it. But for
robustness, the root handle server probably should compare the key to
the PK handle suffix on every operation.

Spot check audit queries are already supported by bind. Additional
code to support audit trails will be very valuable, but not
essential. Reasonable audit archives should be arranged through bind's
logging services.

Although the handle server should check each incoming record to make
sure that the signature corresponds to the handle name, it should not
concern itself in any other way with the source of the record. A
handle owner may delegate the actual transmission of updates to any
other agent. The signature, not the source address, authenticates each
update record. In particular, after an irrevocable transfer of a
handle, the recipient should archive a signed copy of the transfer and
renew it in the root handle server if it gets lost.

The handle server should return additional information very
liberally. As much as possible, it should return signatures even when
they are not requested. Chains of delegation may sometimes get too
long to send in their signed entirety. But for full support of
irrevocable transfer, the handle server must return the transfer
record and its signature, as well as the address that the handle
finally resolves to. This is crucial so that the querier of a
transferred handle may update his own copy of the handle.

\subsection{Reverse Mapping}

Reverse mapping of hosts to handles is not a necessary part of ONHS.
A root handle server should follow the normal current best practice in
reverse mapping for the hosts in its own zones, including those
hosting handle servers. But a root handle server should not be
concerned with reverse mapping of handle owner's hosts.

ONHS users may eventually find it convenient to assign a handle as the
canonical name of each host. If so, each handle owner should take the
usual steps to establish appropriate reverse mapping. But many handles
are likely to be associated temporarily with hosts for reasons that
make reverse mapping to them inappropriate. Even if each
host has a host handle, the owner of host and handle may prefer to use
a conventional domain name as the host's canonical name, and associate
that canonical domain name with the handle by delegation or some other
binding.

\subsection{Handle Operations Imperfectly Supported by DNS Implementation}

\subsubsection{Missing Operations}

DNS/DNSSEC doesn't appear to support signed cancellation directly. \texttt{NXT}
records provided signed information that a given handle is not
currently in use, but that is not at all the same as permanent
cancellation. We might decide to encode cancellation through an
irrevocable binding to a particular address that would never make
sense as a real handle binding (I suggested this method in the example
above). Or, we might have one phoney child handle below each real
handle, and signify cancellation of the real handle by deleting all
children, including the phoney one. I believe that a zone owner may
indicate the complete lack of children by a \texttt{NXT} record with the root
domain on both sides, but I haven't seen a positive verification of
that. A \texttt{NXT} record in the parent will not do, since it must be signed
by the parent's key instead of the child zone's.

The announcement of compromise is not supported directly. It might be
done out of band with a revocation list. It might be simulated with
\texttt{TXT} records (I used this method in the example).

\subsubsection{Weak Support for Irrevocable Operations}

In spite of the usual security reasons to time out all signatures, it
seems best to let signed irrevocable cancellations, transfers, and
compromise markings live indefinitely. The signer of an irrevocable
operation may not be available to resign it, and there seems to be
more value in providing the ancient record and letting the querier
decide its value for himself, than in deleting it. In the case of
transfer, the recipient of a transfer usually has the greatest
interest in making sure that the transfer record is preserved, and the
source of the transfer may even have an interest in repudiating it. In
the case of cancellation and compromise, the owner probably has an
interest in preserving the record, but he may not be able to keep
resigning.

We should assume that every key will eventually be compromised. If the
particular key is not cracked, eventually the whole cryptographic
method is likely to become obsolete due to a combination of
mathematical advances and increases in computing speed. Even so, as
long as anyone is querying it, the best remaining record associated
with a handle has some value. If the value of the handle is high
enough to warrant the trouble, a trusted third party may time-stamp
and sign its final irrevocable record (presumably a transfer). That
signature may be renewed as long as warranted by the value of the
handle. It should usually use stronger cryptography than the owner's
signature.

Because the handle owner may not be able to resign these irrevocable
records, or may not have sufficient incentive to do so, they should be
resigned whenever affordable by a trusted certifier. There is a
serious logistic problem in having a distant trusted certifier resign
such records meaningfully at regular intervals, so it will be helpful
to have an additional signature from the handle root zone, which
presumably will be able to resign regularly. In principle, the handle
root zone should sign the trusted certifier's signature which signs
the handle owner's signature. But DNSSEC doesn't appear to support
signatures on signatures.

Notice that transfer essentially provides a window of time in which
agents may query the source of the transfer and update their links to
use the new handle in place of the old. How long we wish to preserve a
transfer record depends inversely on the frequency of the slowest
querier's queries. We should assume that there are all sorts of
valuable uses of handles other than the ones they are first intended
for, even archaeological sorts of uses. At some point, an old handle
transfer must be moved onto archival storage, but it should be
preserved somewhere as long as possible.

\subsection{Other Types of Resource Records}

When ONHS is implemented on top of DNS, it is straightforward for
handle owners to enter other sorts of signed records for their
handles: \texttt{MX} (mail exchanger), \texttt{HINFO} (host information), \texttt{PTR} (pointer
not followed in resolution), \texttt{RT} (route through), \texttt{TXT} (uninterpreted
text), \texttt{WKS} (well-known services). Since handle owners operate their
own name servers, there is no direct way to prevent them from entering
such records. In principle, the handle root server could detect this
and cut off offending handle servers, but that is probably not
sensible.

In the short run, there is possible benefit, and no harm, from the use
of whatever records DNS allows as values of handles. But, if ONHS is
sufficiently successful, it should probably move to a native
implementation in the future, and such a move should not be burdened
with unfortunate legacy bindings. Records that essentially represent
some sort of generalized address (\texttt{MX} and \texttt{RT}) are likely to be
supported in a future ONHS. With additional protocol agreements at the
edge of the network, \texttt{TXT} records may be used to implement new
experimental sorts of addresses. \texttt{WKS} is a marginal case. In effect it
provides an additional hierarchy of virtual handles below a given
handle. It is probably better to provide such a hierarchy explicitly
in handle values or in some other name hierarchy, such as a new ONHS
zone of DNS. Handle owners probably should eschew \texttt{HINFO} and \texttt{PTR}
records for their handles.

The attachment of NS servers to leaf handles is a special case treated
in Section 3.8.

\subsection{Interleaved Handle Zones and Other Zones}

Conceptually, a leaf handle should resolve to the address of some sort
of agent. In the long run, these addresses should probably be
generalized beyond IP numbers to accommodate agents more loosely
associated with hosts. For example, a future ONHS should probably
accommodate UDP addresses consisting of an IP number and a port
number.

It also makes perfect conceptual sense to associate an ONHS leaf
handle with a DNS name server. (Conceptually, a nonleaf handle is just
a handle assigned to another handle server.) An ONHS leaf handle
associated with a DNS name server is a leaf from the point of view of
ONHS, but not from the point of view of DNS. Such a binding is
implemented by an \texttt{NS} record for the given leaf handle. Since we are
implementing handle servers as a special case of name servers,
leaf-to-name-server resolution allows handle zones to be interleaved
freely with other sorts of DNS zones. Such interleaving appears to be
valuable, and should be encouraged rather than discouraged.

A \texttt{DNAME} record may also be used to assign a name server to a handle,
when the name server already has an independent position in the DNS
name space. The use of \texttt{NS} within a handle zone proposed above provides
a general-purpose DNS name server that is known only through the
handle.

To maintain the conceptual clarity of the restrictions imposed by
ONHS, a handle owner should not mingle publicly advertised textual
domain names as siblings of handles within the same zone. (The syntax
of RRs appears to require domain names for name servers, but those may
be conceived as private names, or even allocated in separate
nonhandle zones.)

If a single public handle service succeeds in supporting the
Internet's need for global handles, then there will be little or no
embedding of other handle servers below textual domain names. The
normal conceptual use of handles has a name space conceptually
separate from handle space, and mapped into it. If that concept
catches on, then extra handle spaces deeper in the domain name
hierarchy will probably be private ones, possibly only visible within
particular intranets. But there is no need to restrict the
interleaving structure. Rather we should experiment to discover the
useful interleavings of handles and names.

\subsection{Possible Future Extensions}

The conceptual foundations of handle systems, and practical
user-interaction issues, suggest a number of possible future
extensions and variations of ONHS. Some of these might be
implementable on future extensions to DNS, or added to a DNS
implementation by judicious interaction with non-DNS software, or
implemented in a native ONHS. All such decisions should be reserved
until early experience provides guidance.

\begin{itemize}
\item Conceptually, a handle is a permanent anchor for a pointer to the
current address of an agent. But not every agent is identified at a
given time with the IP number of a host. IPv6 addressing already
proposes to let IP numbers refer to multicast and anycast groups. But
the ideal network handle system should probably accommodate a greater
variety of agents not identified with hosts, possibly including:
\begin{itemize}
\item subhost agents, identified by a host, an application identifier, and
possibly some parameters to the application (UDP addresses are
examples, where the application is identified with a port; URLs and
email addresses are other examples);
\item distributed agents (multicast and anycast groups are examples, but
there may be more complicated examples);
\item mobile and intermittently connected agents, requiring time-dependent
addresses (I am concerned here with an agent moving between hosts,
rather than mobile hosts which are addressed in IPv6; only an agent
whose address or reachability changes faster than he can send updates
needs ONHS support in this case, others may just keep changing their
handle bindings);
\end{itemize}

When considering whether to accommodate a proposed type of generalized
address, we should consider two key criteria:

\begin{enumerate}
\item whether the generalized address represents a new useful sort of
abstract or virtual agent;
\item whether the service provided by the generalized address can be
simulated effectively and efficiently through direct communication
between a querier and a more primitive type of address.
\end{enumerate}
For example, anycast addresses are desirable because (1) they represent
distributed agents using several equivalent hosts to achieve
reliability through redundancy, and (2) they cannot be simulated
effectively by a single address, since the querier needs to know of
the second address precisely when the first is unreachable.
\item We might wish to offer additional verification services to confirm
important updates, particularly irrevocable ones. For example, we
might delay execution of an irrevocable transfer while notifying the
handle owner through a designated email address and waiting for
confirmation. But every additional layer of verification adds a new
administrative burden and a new exposure to conflict, and it creates a
new vulnerability to denial of service, so we need to consider very
carefully before offering such services. Even optional services create
the vulnerability of fraudulent exercise of the option.
\item As long as handle zones interleave with other domain name zones, a
protocol to distinguish handle zones will be valuable.
\end{itemize}

\section{Choosing a Hash Function}

Since RSA public keys can be engineered to contain specific
substrings, it is important to apply a secure hash function. SHA1
appears to be the current best practice. Individual handle owners may
choose their own tradeoffs between code length and collision
insurance. So each handle key consists of at least 14 hexadecimal
digits from the lower-order end of the 160-bit (40-hexadecimal-digit)
SHA1 hash of an RSA public key. We recommend at least 20 digits from
the hash, and there is often no harm in just using all 40 digits. As
long as we include leading zeroes, the code presents its own length.

In principle, there should be a native implementation of ONHS, with
handles stored in binary. For complete reliable compatibility with the
current DNS, only alphanumeric characters and hyphens are safe, and
even case-dependence seems risky. So base 64 is slightly out of
reach. Base 32 is feasible, but base 16 (hexadecimal) is only 5/4
longer and it's the usual way to present a hash key. I also considered
base 10 presentation of the hash code. But the length of base 10
numbers doesn't correspond as cleanly with the length of binary
numbers.

\section{ONHS on DNS with IPv6}

The implementation of ONHS in DNS depends on DNAME resource records,
which are associated with the transition to IPv6, but do not depend on
it. Without DNAME support, CNAME allows delegation only of leaf
handles.

Since ONHS doesn't manipulate IP addresses except to store them when
assigned to handles and return them unchanged in response to
resolution queries, upgrade of DNS to IPv6 automatically upgrades
ONHS. IPv6 handle servers merely store AAAA or A6 records instead of A
records.

\section{Comparison to Related Systems}

\subsection{Domain Name System}

The Domain Name System (DNS) appears to have been designed primarily
as a system to provide permanent handles that can be reassigned to
different addresses to accommodate slow mobility of hosts,
reassignment of functions to different hosts, and changes in
addressing dictated by changes in network topology and routing
efficiency. It was designed before public-key techniques were widely
deployed, and used textual names to make handles easier to remember,
type, and communicate out of band. As a result, DNS today is known
largely as a particular distributed hierarchical directory of
names. The meaning relationships in the name hierarchy compete with
the pure ownership authority relationships, and arguably dominate them
at least near the root [DNS].

ONHS is essentially a restriction of DNS service to meaningless
numerical handles, abandoning support of the convenience and value of
meaningful names to other related services, such as DNS. By supporting
only the continuity of handles, without the human meaning of names,
ONHS hopes to derive two advantages:
\begin{itemize}
\item self-assignment of handles through public-key techniques;
\item avoidance of conflict over the meanings of names.
\end{itemize}

Although a handle, by itself, is inherently less valuable than a name
with a handle, by unbundling we should be able to provide handles more
promiscuously, efficiently, and cheaply, and to free them from the
competition-generated scarcity and conflict associated with names.

\subsubsection{Security Extensions}

ONHS can take advantage of public-key cryptographic functionality in
the DNS security extensions [DNSSEC]. But the intention and expected
effect of the use of cryptographic techniques in ONHS is quite
different from DNSSEC.

Signatures in DNSSEC are intended to authenticate communications among
name servers and between name servers and resolvers. The authority for
a particular record is invested in a name server. A signature with
that server's key insures that the particular name server is
the true source of the record. The association of a public key with
the identity of a name server is itself signed by a higher authority,
using a chain of trust up to some security root authority whose key is
distributed reliably out of band.

Authority in ONHS is invested directly in the public keys
themselves. No particular handle server is invested with authority as a
handle server. Of course, keys and handle servers are likely to be
co-owned, but ONHS takes no particular interest in, nor responsibility
for authenticating, this co-ownership.

Handle servers are responsible for best effort resolution of handles
to addresses, but are not responsible for the correctness of
individual resolutions. Anyone who is dissatisfied with the
performance of existing handle servers may operate, or contract with a
third party for the operation of, an additional server to improve the
rate of correct resolutions. A rogue handle server may cause denial of
service by flooding, but by itself it may not cause a querier to use a
fraudulent resolution.

The hierarchical structure of handle space allows construction of
chains of trust, but these are not normally used with public-key
handles. Rather, each public-key handle is normally thought of as an
independent security root. ONHS creates no trust in any handle, but it
allows individual transactions providing trust through out-of-band
mechanisms to accumulate reliably around a handle (or even a transfer
chain of handles). The chain of trust approach may be used to
incorporate authentication techniques weaker than public-key
techniques, such as password-authenticated handle updates.

\subsection{Uniform Resource Names}

The Internet Engineering Task Force (IETF) has a working group on
Uniform Resource Names (URN). URNs are "persistent identifiers for
information resources." At the motivational level, the mission of URNs
appears very similar to that of ONHS handles. But the URN working
group has concerned itself very much with the semantic relationships
between URNs imposed by various authorities. To my knowledge they have
not proposed any sort of self-assigned URNs [URN].

A successful implementation of ONHS handles could provide the
persistence required of URNs, allowing the URN project to focus more
on additional services to establish semantic relationships.

\subsection{Uniform Resource Identifiers}

A related working group is concerned with Uniform Resource Identifiers
(URI). A URI is "a compact string of characters for identifying an
abstract or physical resource." URIs appear to be intended as a
broader class of objects, including URNs as well as less persistent
identifiers [URI]. I expect that URIs will resolve to URNs, but I'm
not sure I've understood the working groups' intentions correctly. The
URI working group appears to be particularly concerned with human
readability, which is not at all a concern of ONHS.

\subsection{Simple Public Key Infrastructure; Simple Distributed Security Infrastructure}

A working group on Simple Public Key Infrastructure (SPKI) merged with
Ronald Rivest's and Butler Lampson's Simple Distributed Security
Infrastructure (SDSI) project. A key component of SPKI/SDSI formalizes
the use of names across different naming contexts. Whenever Sally uses
the name "Paul" for one agent, who uses the name "George" for another
agent, we may refer to "Sally's Paul's George." Chains of names are
rooted in self-assigned public-key names [SPKI]. ONHS is essentially a
hierarchical specialization of the SPKI/SDSI naming system, limiting
the use of names so that they resolve only to network addresses.

\subsection{Open Privacy Initiative's Nyms}

ONHS's basic idea of public keys as handles, allowing an accumulation
of trust, is the same idea already used by the Open Privacy Initiative
(OPI) in its Nyms. Nyms are public-key handles, with additional
capabilities. For example, the owner of several nyms may prove their
relation at will, or keep it private [OPI-Nym]. So far, I haven't seen
the need for that service in ONHS. I hope that it may be added to
basic ONHS service by a separate service, when an application demands
it. OPI also provides tools to accumulate good and bad reputation
around a Nym. ONHS provides handles as anchors for such accumulation,
but does not support the accumulation itself. Essentially, ONHS is the
application of the Nym idea purely to network addressing, leaving all
other useful functions with Nyms to add-on services.

\section{References}

\subsection{Normative References}

\subsection{Informative References}

\section{Security Considerations}

ONHS doesn't appear to generate any substantial new security risks. It
is vulnerable to the same sorts of attacks as other uses of DNS. If
the system is not advertised clearly and honestly, users might depend
on it for verification of identity, which it does not provide, leading
to attacks on those users based on their credulity.

I believe that a root handle zone may be operated as a fully compliant
secure zone. But it must mark all of the subzones for individual
handle owners as insecure, so knowing the security of a root handle
zone is not very helpful. And the actual security value of the root
handle zone signatures is very small.

Although ONHS is implementable under DNS, the totally distributed
authentication model in the conceptual foundations of ONHS does not
match the chain-of-trust model in the foundations of the DNS security
extensions. It is probably best to just treat all handle zones as
insecure from the point of view of DNS.

\subsection{Confidentiality}

ONHS is not at all concerned with confidentiality. ONHS data are
completely public. Agents desiring confidentiality must find other
means to achieve it. Encryption of ONHS messages might be used as a
defense against man-in-the-middle attacks, but it should not be used
for basic confidentiality, since any Internet user may query the
entire handle data.

\subsection{Data Integrity}

The ONHS root handle server is not directly and essentially concerned
with the data integrity of handle bindings. Each querier may verify
the authenticity of signed bindings for himself. But a failure of data
integrity within the system may damage the performance of the ONHS
system, either flooding it with inauthentic messages or even causing
authentic data to be ignored or discarded. Each handle owner may
protect against permanent loss of data by keeping her own archive. So
the real problem is that internal failure of data integrity leads to
denial of service.

There is no complete defense against denial of service by flooding at
the level of ONHS. By verifying each signature before storing a record
in its database, a handle server limits the transmission of
inauthentic packets and reduces, but does not eliminate, the impact of
a flooding attack.

\subsection{Peer Entity Authentication}

Like data integrity, peer entity authentication is not absolutely
crucial to the internal operations of ONHS. Each record is
recognizably authentic or inauthentic on its own. An impostor name
server who transmits authentic records does little or no harm, as long
as resolvers and queriers avoid drawing conclusions beyond the
official meanings of the records.

On the other hand, ONHS provides a tool that may help authentication
of the handle owner by the querier, directly by connecting the querier
to an address provided on the authority of the handle owner, and
incidentally by providing a public key that the querier and handle
owner may choose to use for further authentication. But ONHS does not
provide any sort of assurance of the identity or other quality of the
handle owner. Such assurance must be derived from communications with
the address provided by the handle owner, or through other services
that link to handle space in some way.

\subsection{Nonrepudiation}

Nonrepudiation is not very important to ONHS operations. ONHS provides
connections allowing queriers and handle owners to communicate. All
qualities affecting the two agents' satisfaction should be determined
by the content of that communication and/or other services outside of
ONHS.

For auditing purposes, it is helpful that the signatures on bindings
and certain negative results (\texttt{NXT}) are hard to repudiate by the handle
owner, but the inability to identify the handle owner limits the value
of that nonrepudiation. It is also helpful that the handle server's
signatures on its metadata are hard to repudiate. This nonrepudiation
is mostly useful for voluntary debugging of the system, since a handle
server should not provide strong warranties of service.

The nonrepudiation quality of ONHS operations is only at the level of
public-key signature, and therefore only valuable under the assumption
that the key has not been discovered by an adversary, and that no
adversary has tricked an agent into signing the wrong record. But the
potential damage due to repudiation is also light. ONHS only takes
responsibility for carrying out resolution according to correctly
signed records. Each handle owner bears the consequences of her own
competence or incompetence in key management.

\subsection{Systems Security, Unauthorized and Inappropriate Usage}

ONHS implemented on DNS has little or no impact on systems
security. Vulnerabilities are essentially those of the DNS bind
software. There is a small additional exposure of the handle root name
server due to the fact that it communicates subzone data with unknown
parties, but as long as it limits that communication to accepting
signed resource records it should be easy to prevent attacks through
such communication.

Perhaps the most serious security problems introduced by ONHS will
come from attacks on queriers who accept unsubstantiated claims of
authenticity based merely on ONHS's response to queries. To defend
against those attacks, we should educate end users and those who
provide them with handle-aware software to present the meaning of
handle resolution honestly, accurately, and understandably. These
attacks on queriers may also harm handle owners who depend too
strongly on ONHS to guarantee connections and thereby lose important
correspondents. A handle owner may also be harmed if she is blamed for
behavior of an attacker who hijacks the handle. A service that
connects a handle to some other information about reputation or
identity should take its own steps to defend against the consequences
of handle hijacking.

The ability to notify ONHS of a compromised password and cancel it
irrevocably is a final defense for a handle owner against unauthorized
use of the handle. The use of irrevocable cancellation reduces the
consequences of unauthorized use to denial of service. The seriousness
of this denial depends on the value of the connections with queriers
enabled by the handle, and the difficulty of re-establishing those
connections out of band. The handle owner whose handle is particularly
valuable should use longer keys and stronger key-management techniques
proportional to the potential damage.

A client of a password-authenticated handle service is vulnerable to
sniffed or cracked passwords. But standard SSL and/or ssh techniques
may be used to avoid passing passwords in plaintext. The owner of a
password-authenticated handle may not be able to use compromise
announcement and cancellation as a last defense, since an attacker may
change the password very quickly. The consequences of password change
may be reduced by always implementing password change as irrevocable
transfer to a new handle with the new password. But password
authentication is so inherently weak that we should probably
concentrate on making sure that all handles valuable enough to be
worth a serious cracking effort are protected by stronger techniques
instead of on minor improvements to the password mechanism.

The provider of password-authenticated passwords may defend against
password sniffing/cracking attacks by confirming updates through a
predetermined email address. But every additional layer of
confirmation provides another potential point for denial of service
attack, so such layers should not be added without careful
consideration.

\subsection{Denial of Service}

Denial of service by flooding is probably the threat to ONHS
operations that is hardest to defend against. ONHS exposure is
probably essentially the same as general DNS exposure, whether ONHS is
implemented on DNS or independently. ONHS may suffer somewhat greater
exposure, since it inherently must communicate with unknown handle
owners. But the loose connection works both ways. Each handle owner is
responsible for her own handle server. ONHS's obligation to an
individual handle owner is light. A handle root server may
systematically refuse communication from an IP address that has been
flooding it, with relatively small service consequences from the
refusal.

An adversary may use ONHS to try to direct innocent traffic to an
inappropriate address as part of a denial of service by flooding
attack on that address. This is essentially the same as the "slashdot
effect." The adversary may obtain the handle legitimately, but he must
advertise some attractive quality for the handle outside of the
system. Since the adversary could just as well advertise such a
quality for the IP address, or for a domain name, the added exposure
due to ONHS is small or nil.

Defense against denial of service by flooding using ONHS as an
intermediary to channel innocent traffic to an inappropriate address
should be defended mostly by counteradvertising in the same or similar
channel to that used by the attacker. In an extreme case, a handle
server administrator may cease resolving a given handle on the well
authenticated objection of the owner of the address to which the
handle has been assigned. But that step should be taken with care,
since it opens up a new line of denial of service attack by an
adversary posing as the owner of the address. When IPv6 is widely
deployed, the victim of such a flooding attack can probably defend by
changing the IP address.

\subsection{Types of Attack}

\subsubsection{Eavesdropping}

ONHS is immune to eavesdropping by itself, since all of its data are
public anyway. An eavesdropper only learns what he might have learned
by querying the system, or through a procsy who queries the
system. Eavesdropping may be a component of a man-in-the-middle
attack, but eavesdropping alone does no harm. The best defense against
eavesdropping as a component of another attack is probably encryption
of ONHS messages.

\subsubsection{Message Replay}

ONHS defends against message replay in the same way that DNS does. It
follows the principle that every message in a distributed system must
carry its whole meaning internally. So a message replayed out of
sequence does not change the operation. This property is sometimes
referred to as commutativity plus idempotence. Final results depend
only on the set of transactions, not on the order or multiplicity. The
serial number or time stamp in each update is a key part of this
defense: without it replay could reset a handle binding to a previous,
now invalid binding. Replay might enhance a denial of service by
flooding attack, by consuming more system resources before a message
can be discarded.

ONHS carries the principle of complete context-free meaning for each
message even farther than normal operations of DNS with security
extensions. DNSSEC applies this principle to the database content of
messages, before the signature, but it fails to follow the principle
with respect to signatures. The meaning of a signature in DNSSEC
depends not only on the signature itself, but also on the chain of
trust by which the signature is attached to an authority. In ONHS, the
signature is bound to a handle by the value of the handle itself, so
the value of a signature is much less context-dependent. There's no
magic here: ONHS makes a much weaker claim for the value of a
signature since it does not verify the connection between the
signature key and any independently identified agent.

This extra context-independence in ONHS doesn't make it invulnerable
to any particular type of attack to which DNSSEC is vulnerable. It
just reduces the number of points of failure to one for a public-key
handle, while DNSSEC has a point of failure for each link in the chain
of trust. And, to repeat and re-emphasize, it reduces vulnerability
while also reducing the strength of its claims for its results.

\subsubsection{Message Insertion}

ONHS defends against message insertion by signing every message. A
message insertion attack requires a compromise of a handle owner's
key. Once the key is compromised, an adversary may enter arbitrary
fraudulent data in that handle's records. As long as the owner does
not lose the key, she may still reduce the damage to denial of service
by announcing compromise and canceling the handle. A quick-acting
adversary may transfer the handle, but compromise announcements are
recorded even after transfer, and reported to all who query the old
handle.

An adversary who transfers a handle captures traffic from those who
query the handle before the compromise announcement and replace their
links with the new handle. A handle server may defend against this
sort of hijacking by reporting the compromise announcement to queriers
of the target handle as well. But that step should be taken only after
careful consideration and out-of-band authentication, since it
provides a new way to attack the value of one handle by transferring
another to it, then declaring a compromise. This new sort of attack
can be carried out with a handle acquired legitimately by the
attacker, so it is quite easy. A service to report transitive
compromises should probably be implemented as a separate registry
outside of ONHS, with very carefully thought out authentication
methods.

The harm of the message insertion attack may be increased by a man in
the middle attack (Section 8.7.6) that delays the handle owner's
discovery of the fraud.

\subsubsection{Message Deletion}

ONHS is vulnerable to message deletion. The main harm of message
deletion is denial of service. Message deletion by itself does not
yield fraudulent results. An attacker may use message deletion to
prolong a handle-address binding beyond its validity. Then, if the
attacker compromises the old address, he might hijack some
traffic. But as long as the relevant key is not compromised, each
querier may discover the fraud, reducing the harm to denial of
service. Harm may be reduced further by the handle owner's use of
relatively short expiration dates on bindings, but that also only
reduces the harm from message hijacking to denial of service.

More positive defenses against message deletion are multiple redundant
communication paths, through multiple redundant handle servers, and
confirmation with resend, so that the attacker must accomplish several
co-ordinated deletions. A handle owner should verify every important
update by a test query to confirm the effect.

Eavesdropping plus message deletion plus insertion may also be used to
thwart a defense against fraudulent update that uses confirmation,
e.g. by email, particularly if the confirmation uses less secure
authentication (which it is likely to do in the case of email
confirmation). This is almost a man in the middle attack, but it
only requires control of the handle server's end of the channel, not
the handle owner's end.

The harm of message deletion attacks may be increased by a man in the
middle attack (Section 8.7.6) that delays the handle owner's discovery
of the fraud.

\subsubsection{Message Modification}

ONHS vulnerability to message modification is essentially the same as
to message insertion. An attacker who discovers a secret key may send
any sort of message just as easily as modifying a legitimate
message. Modification instead of construction from scratch may allow
the attacker to make the fraudulent message more credible to
out-of-band auditing. But essentially, modification is not a problem
unless the key is compromised, and then the handle becomes worthless
and potentially harmful to legitimate users.

\subsubsection{Man in the Middle}

Basic ONHS operations are not vulnerable to man in the middle attacks,
since each transaction is atomic, except of course when the man in the
middle has discovered a private key. Once a handle owner's private key
is compromised, a man in the middle may delay the handle owner's
discovery of the compromise indefinitely, by returning correct results
to all test queries by the handle owner while sending incorrect
updates to the handle server. The best defense against this sort of
attack is probably multiple communication paths through multiple
handle servers, so that the man in the middle must cover a broader
middle. In particular, a handle owner worried about such an attack
should make covert queries through procsies to verify correct updates
to her handle records.

A man in the middle may also thwart the defense against fraudulent
updates using reconfirmation through alternate channels. In this case
the man in the middle may either provide no information at all to the
handle owner, or he may give the handle owner a false impression that
she has thwarted a fraudulent transaction. This is probably mostly a
threat against password-authenticated handles, since the owner of a
public-key handle should immediately announce a compromise and
cancellation after thwarting a fraudulent update. If the man in the
middle can intercept and discard those announcements, but produce
confirmation of them, he can delay discovery of the harm indefinitely.

The handle owner's best defense is probably multiple channels of
communication to broaden the coverage required of the attacker. The
handle owner and handle server may agree to defend by also requiring
multiple confirmations through different channels. But each additional
confirmation is an additional point of attack for denial of service.

Encryption of traffic, even though the traffic itself need not be
confidential, defends against the man in the middle by giving him one
more secret key to discover.

A man in the middle may also destroy or falsify audit trails. Handle
owners and other auditors may defend with periodic spot check audit
queries and appeals to the archival logs, through multiple channels to
multiple handle servers.

There is essentially no defense against a man in the middle with
sufficient power. If the man in the middle controls all channels of
communication to any agent, and has discovered all secret keys, he can
create a complete false reality, and trick the agent into actions that
void any sort of mathematical security. We cross our fingers that our
adversaries will never be quite that powerful.

\section{ANA Considerations}

The proposed DNS implementation of ONHS uses the IANA DNS Security
Algorithm Numbers. Since the cryptographic concerns of DNS and ONHS
are very similar, I expect that future updates to these numbers will
continue to support both services. I don't anticipate any demands on
IANA from ONHS in the foreseeable future. If an ONHS implementation on
DNS is sufficiently popular, and if there are many different handle
zones, then we may want some way to distinguish handle zones from
other DNS zones, and that may call for an assignment of a code by
IANA.

\section{Acknowledgments}

Bob Frankston pointed out the value of a safe haven within the domain
name space providing permanent cheap handles not subject to trade name
disputes (\texttt{http://www.frankston.com/public/ESSAYS/DNSSafeHaven.asp}).
When I was pondering how to use asymmetric cryptographic techniques
for self-assigned handles, Scott Nelson suggested hashed public keys,
and gave me a quick education in the structure of RSA public keys. I
later discovered that Daniel J.\ Bernstein had mentioned the value of
host names that match public keys
(\texttt{http://cr.yp.to/djbdns/forgery.html}). The Open Privacy Initiative
proposes "nyms," which use a similar idea but make the connection
between sibling handles private until revealed by the owner\\
(\texttt{http://www.openprivacy.org/papers/200103-white.html}).
Carl M.\ Ellison
also mentioned the use of public keys and their hashes as
self-assigned unique identifiers
(\texttt{http://world.std.com/~cme/usenix.html},\\
\texttt{http://www.cfp2000.org/papers/ellison.pdf}). Rivest and Lampson use the
same basic idea in SDSI\\
(\texttt{http://theory.lcs.mit.edu/~cis/sdsi.html}).

\section{Author's Address}

\begin{trivlist}
\item Michael J. O'Donnell
\item The University of Chicago
\item Department of Computer Science
\item Ryerson Hall, 1100 E. 58th St.
\item Chicago, IL 60637
\item USA
\item Phone: 773 702-1269
\item FAX: 773 702-8487
\item email: \texttt{michael\_odonnell@acm.org}
\end{trivlist}

\section{Full Copyright Statement}

Copyright held by the author, Michael J. O'Donnell (2002). I transfer
copyright to The Internet Society on the sole condition that they
choose to publish this article as an Internet Draft. In that case,
this copyright statement is to be replaced by the following:

\begin{quote}
"Copyright (C) The Internet Society (2002).  All Rights Reserved.

This document and translations of it may be copied and furnished to
others, and derivative works that comment on or otherwise explain it
or assist in its implementation may be prepared, copied, published
and distributed, in whole or in part, without restriction of any
kind, provided that the above copyright notice and this paragraph are
included on all such copies and derivative works.  However, this
document itself may not be modified in any way, such as by removing
the copyright notice or references to the Internet Society or other
Internet organizations, except as needed for the purpose of
developing Internet standards in which case the procedures for
copyrights defined in the Internet Standards process must be
followed, or as required to translate it into languages other than
English.

The limited permissions granted above are perpetual and will not be
revoked by the Internet Society or its successors or assigns.

This document and the information contained herein is provided on an
"AS IS" basis and THE INTERNET SOCIETY AND THE INTERNET ENGINEERING
TASK FORCE DISCLAIMS ALL WARRANTIES, EXPRESS OR IMPLIED, INCLUDING
BUT NOT LIMITED TO ANY WARRANTY THAT THE USE OF THE INFORMATION
HEREIN WILL NOT INFRINGE ANY RIGHTS OR ANY IMPLIED WARRANTIES OF
MERCHANTABILITY OR FITNESS FOR A PARTICULAR PURPOSE." "
\end{quote}

\end{document}